\documentclass[twocolumn,english,journal]{IEEEtran}
\usepackage[T1]{fontenc}
\usepackage{color}
\usepackage{babel}
\usepackage{amsmath}
\usepackage{amssymb}
\usepackage{graphicx}
\usepackage[unicode=true,
 bookmarks=true,bookmarksnumbered=true,bookmarksopen=true,bookmarksopenlevel=1,
 breaklinks=false,pdfborder={0 0 0},backref=false,colorlinks=false]
 {hyperref}
\hypersetup{pdftitle={Your Title},
 pdfauthor={Your Name},
 pdfpagelayout=OneColumn, pdfnewwindow=true, pdfstartview=XYZ, plainpages=false}

\makeatletter
\usepackage[caption=false,font=footnotesize]{subfig}

\newtheorem{assumption}{Assumption}
\newtheorem{theorem}{Theorem}
\newtheorem{remark}{Remark}
\newtheorem{definition}{Definition}

\makeatother

\begin{document}

\title{Data-driven controller design for nonlinear systems:\\
a two degrees of freedom architecture}

\author{Carlo Novara and Simone Formentin
\thanks{Carlo Novara is with Dipartimento di Automatica e Informatica, Politecnico di Torino, Italy, e-mail: \protect\href{http://carlo.novara@polito.it}{carlo.novara@polito.it}. Simone Formentin is with Dipartimento di Elettronica, Informazione e Bioingegneria, Politecnico di Milano, Italy, e-mail: \protect\href{http://simone.formentin@polimi.it}{simone.formentin@polimi.it}.} 
}

\maketitle
\begin{abstract}
In this paper, the D$^{2}$-IBC (Data-Driven Inversion Based Control) approach for nonlinear control is introduced and analyzed. The method does not require any a-priori knowledge of the system dynamics and relies on a two degrees of freedom scheme, with a nonlinear controller and a linear controller running in parallel. In particular, the former is devoted to stabilize the system around a trajectory of interest, whereas the latter is used to boost the closed-loop performance. The paper also presents a thorough stability and performance analysis of the closed-loop system.
\end{abstract}

\section{Introduction}

\label{sec:ibc_approach}

Consider a nonlinear discrete-time SISO system in regression form:
\begin{equation}
y_{t+1}=g\left(\boldsymbol{y}_{t},\boldsymbol{u}_{t},\boldsymbol{\xi}_{t}\right)\label{ss_sys}
\end{equation}
\[
\begin{array}[t]{c}
\boldsymbol{y}_{t}=\left(y_{t},\ldots,y_{t-n+1}\right)\\
\boldsymbol{u}_{t}=\left(u_{t},\ldots,u_{t-n+1}\right)\\
\boldsymbol{\xi}_{t}=\left(\xi_{t},\ldots,\xi_{t-n+1}\right)
\end{array}
\]
where $u_{t}\in U\subset\mathbb{R}$ is the input, $y_{t}\in\mathbb{R}$
is the output,$\xi_{t}\in\Xi\subset\mathbb{R}^{n_{\xi}}$ is a disturbance
including both process and measurement noises, and $n$ is the system
order. $U$ and $\Xi$ are compact sets. In particular, $U\doteq[\underline{u},\overline{u}]$
accounts for input saturation.

Suppose that the system (\ref{ss_sys}) is unknown, but a set of measurements
is available:
\begin{equation}
\mathcal{D}\doteq\left\{ \tilde{u}_{t},\tilde{y}_{t}\right\} _{t=1-L}^{0}\label{eq:data}
\end{equation}
where $\tilde{u}_{t}$ and $\tilde{y}_{t}$ are bounded for all $t=1-L,\ldots,0$.
The accent $\sim$ is used to indicate the input and output samples of the
data set \eqref{eq:data}.

Let $\mathcal{Y}^{0}\subseteq\mathbb{R}^{n}$ be a set of initial
conditions of interest for the system (\ref{ss_sys}) and, for a given
initial condition $\boldsymbol{y}_{0}\in\mathcal{Y}^{0}$, let $\mathcal{Y}\left(\boldsymbol{y}_{0}\right)\subseteq\ell_{\infty}$
be a set of output sequences of interest. 

The aim is to control the system (\ref{ss_sys}) in such a way that,
starting from any initial condition $\boldsymbol{y}_{0}\in\mathcal{Y}^{0}$,
the system output sequence $\boldsymbol{y}=(y_{1},y_{2},\ldots)$
tracks any reference sequence $\boldsymbol{r}=(r_{1},r_{2},\ldots)\in\mathcal{Y}\left(\boldsymbol{y}_{0}\right)$.
The set of all solutions of interest is defined as $\mathcal{Y}\doteq\left\{ \mathcal{Y}\left(\boldsymbol{y}_{0}\right):\boldsymbol{y}_{0}\in\mathcal{Y}^{0}\right\} $.
The set of all possible disturbance sequences is defined as $\varXi\doteq\left\{ \boldsymbol{\xi}=(\xi_{1},\xi_{2},\ldots):\xi_{t}\in\Xi,\forall t\right\} $.

\begin{figure}[h]
\begin{centering}
\includegraphics[bb=120bp 410bp 560bp 700bp,clip,scale=0.5]{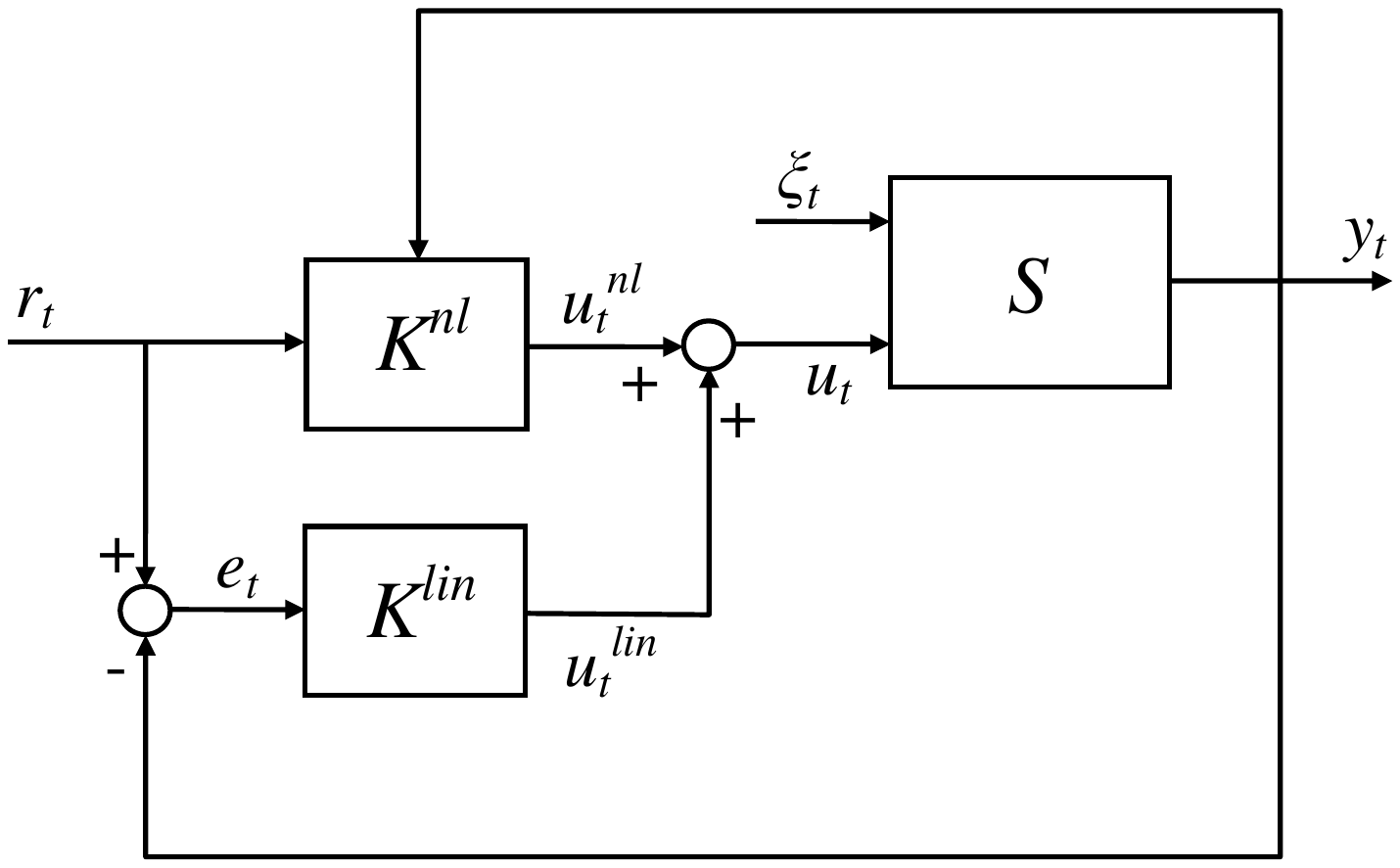} 
\par\end{centering}

\protect\caption{Feedback control system.}

\label{ctr_sys_1} 
\end{figure}

To accomplish this task, we use the feedback control structure depicted
in Figure \ref{ctr_sys_1}, where $S$ is the system (\ref{ss_sys}),
$K^{nl}$ is a nonlinear controller, $K^{lin}$ is a linear controller,
$r_{t}\in Y$ is the reference, and $Y\subset\mathbb{R}$ is a compact
set where the output sequences of interest lie.

$K^{nl}$ is used to stabilize the system (\ref{ss_sys}) around the
trajectories of interest, while $K^{lin}$ allows us to further reduce
the tracking error (especially in steady-state conditions). $K^{nl}$
is designed through the NIC (Nonlinear Inversion Control) approach
presented in \cite{arxiv1}, $K^{lin}$ is designed using a suitably modified version of the VRFT (Virtual
Reference Feedback Tuning) method introduced in \cite{campi2002virtual}. As
shown in Sections \ref{sub:nl_des} and \ref{sub:lin_des}, the design
of both the controller is performed from data and is based on system
inversion, hence the name D$^{2}$-IBC (Data-Driven Inversion Based
Control). 

Besides control design, other main contributions of the paper are
a closed-loop stability analysis and a study on the performance enhancement
given by the linear controller.

\section{Notation}

A column vector $x\in\mathbb{R}^{n_{x}\times1}$ is denoted as $x=\left(x_{1},\ldots,x_{n_{x}}\right)$.
A row vector $x\in\mathbb{R}^{1\times n_{x}}$ is denoted as $x=\left[x_{1},\ldots,x_{n_{x}}\right]=\left(x_{1},\ldots,x_{n_{x}}\right)^{\top}$,
where $\top$ indicates the transpose.

A discrete-time signal (i.e. a sequence of vectors) is denoted with
the bold style: $\boldsymbol{x}=(x_{1},x_{2},\ldots)$, where $x_{t}\in\mathbb{R}^{n_{x}\times1}$
and $t=1,2,\ldots$ indicates the discrete time; $x_{i,t}$ is the
$i$th component of the signal $\boldsymbol{x}$ at time $t$.

A regressor, i.e. a vector that, at time $t$, contains $n$ present
and past values of a variable, is indicated with the bold style and
the time index: $\boldsymbol{x}_{t}=\left(x_{t},\ldots,x_{t-n+1}\right)$.

The $\ell_{p}$ norms of a vector $x=\left(x_{1},\ldots,x_{n_{x}}\right)$
are defined as
\[
\left\Vert x\right\Vert _{p}\doteq\begin{cases}
\left(\sum_{i=1}^{n_{x}}\left|x_{i}\right|^{p}\right)^{\frac{1}{p}}, & p<\infty,\\
\max_{i}\left|x_{i}\right|, & p=\infty.
\end{cases}
\]
The $\ell_{\infty}$ norm is also used to denote the absolute value
of a scalar: $\left\Vert x\right\Vert _{\infty}\equiv\left|x\right|$
for $x\in\mathbb{R}$. 

The $\ell_{p}$ norms of a signal $\boldsymbol{x}=(x_{1},x_{2},\ldots)$
are defined as
\[
\left\Vert \boldsymbol{x}\right\Vert _{p}\doteq\begin{cases}
\left(\sum_{t=1}^{\infty}\sum_{i=1}^{n_{x}}\left|x_{i,t}\right|^{p}\right)^{\frac{1}{p}}, & p<\infty,\\
\max_{i,t}\left|x_{i,t}\right|, & p=\infty,
\end{cases}
\]
where $x_{i,t}$ is the $i$th component of the signal $\boldsymbol{x}$
at time $t$. These norms give rise to the well-known $\ell_{p}$
Banach spaces.

\section{Nonlinear controller design}

\label{sub:nl_des}

The nonlinear controller design is based on the method presented in
\cite{arxiv1}. The first step of this method is to identify from the
data \eqref{eq:data} a model for the system \eqref{ss_sys} of the
form
\begin{equation}
\begin{array}[t]{l}
\hat{y}_{t+1}=f\left(\boldsymbol{y}_{t},\boldsymbol{u}_{t}\right)\equiv f\left(\boldsymbol{q}_{t},u_{t}\right)\\
\boldsymbol{q}_{t}=\left(y_{t},\ldots,y_{t-n+1},u_{t-1},\ldots,u_{t-n+1}\right)
\end{array}\label{eq:model}
\end{equation}
where $u_{t}$ and $y_{t}$ are the system input and output, and $\hat{y}_{t}$
is the model output. For simplicity, the model is supposed of the
same order as the system but this choice is not necessary: all the
results presented in the paper hold also when the model and system
orders are different. Suitable algorithms for model identification
can be found in \cite{ifac14_3} or \cite{arxiv1}.

Once a model of the form \eqref{eq:model} has been identified, the
command action $u_{t}^{nl}$ of the controller $K^{nl}$ is obtained
by the on-line inversion of this model. In the NIC approach, the following
optimization problem is solved to perform such an inversion:
\begin{equation}
\begin{array}[t]{ccl}
u_{t}^{nl} & = & \arg\min_{\mathfrak{u}\in U}J\left(\mathfrak{u}\right)\\
 &  & \textrm{subject to}\;\;\mathfrak{u}\in U.
\end{array}\label{eq:opt2}
\end{equation}
The objective function is given by
\begin{equation}
J\left(\mathfrak{u}\right)=\frac{1}{\rho_{y}}\left(r_{t+1}-f\left(\boldsymbol{q}_{t},\mathfrak{u}\right)\right)^{2}+\frac{\mu}{\rho_{u}}\mathfrak{u}^{2}\label{eq:objf2}
\end{equation}
where $\rho_{y}\doteq\left\Vert \left(\tilde{y}_{1-L},\ldots,\tilde{y}_{0}\right)\right\Vert _{2}^{2}$
and $\rho_{u}\doteq\left\Vert \left(\tilde{u}_{1-L},\ldots,\tilde{u}_{0}\right)\right\Vert _{2}^{2}$
are normalization constants computed from the data set \eqref{eq:data},
and $\mu\geq0$ is a design parameter, allowing us to determine the
trade-off between tracking precision and command activity. This inversion
technique is similar to the one in \cite{NCMS14}, where a Set Memebrship
model is used.

Note that the objective function \eqref{eq:objf2} is in general non-convex.
Moreover, the optimization problem \eqref{eq:opt2} has to be solved
on-line, and this may require a long time compared to the sampling
time used in the application of interest. In order to overcome these
two relevant problems, the technique presented in \cite{arxiv1} can
be used, allowing a very efficient computation of the optimal command
input $u_{t}^{nl}$.

\section{Linear controller design}

\label{sub:lin_des}

The linear controller $K^{lin}$ is defined
by the extended PID (Proportional Integral Derivative) control law
\begin{equation}
u_{t}^{lin}(\theta)=u_{t-1}^{lin}(\theta)+\sum_{i=0}^{n_{\theta}}\theta_{i}e_{t-i}\label{Klin}
\end{equation}
where $e_{t}=r_{t}-y_{t}$ is the tracking error, $n_{\theta}$ is
the controller order and the $\theta_{i}$'s denote the controller
parameters. The goal of $K^{lin}$ in the proposed architecture is to compensate for model-inversion
errors and boost the control performance by assigning a desired dynamics
to the resulting nonlinearly-compensated system.

The Virtual Reference Feedback Tuning (VRFT) method \cite{campi2002virtual,ifac14_0}
is here suitably adapted to be applicable in the D$^{2}$-IBC setting and employed to design the linear controller.

\begin{figure}[h!tb]
\centering \includegraphics[width=0.8\columnwidth]{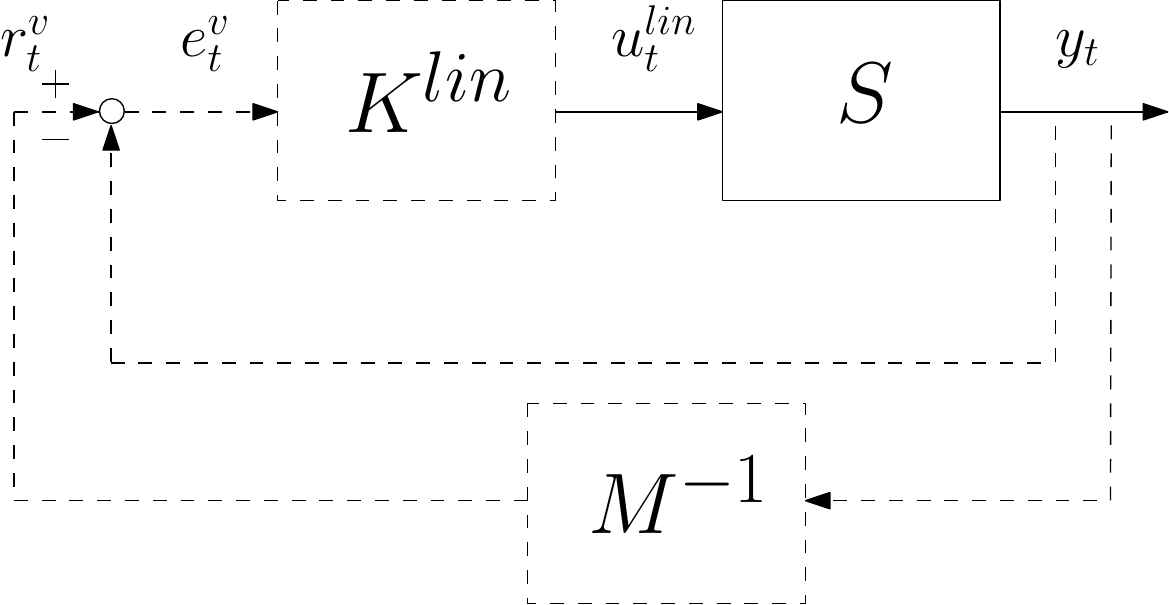}
\protect\protect\caption{The ``virtual reference'' rationale: the data is collected on the
real system (solid) and applied to controller identification in the
``virtual loop'' (dashed).}

\label{fig:vr_idea} 
\end{figure}

Let the desired behavior for the closed-loop system be given by a linear asymptotically stable model $M$.

The ``virtual reference'' rationale to design $K^{lin}$ achieving
$M$ without identifying any model of the system is based on the following
observation, illustrated in Figure \ref{fig:vr_idea}. In a ``virtual''
operating condition where the closed-loop system behaves exactly as
$M$, the ``virtual reference'' signal $r_{t}^{v}$ would be given as the output of the inverse of $M$, say $M^{-1}$, when it is fed by $y_t$.

Obviously, since $M^{-1}$ is likely to be non-causal, $r_{t}^{v}$
could be computed only off-line using the available data set. However,
in such a setting, both the trajectory of the (fictitious signal)
$r_{t}^{v}$ and the subsequent ``virtual error'' $e_{t}^{v}=r_{t}^{v}-y_{t}$
could be calculated. This fact means that the optimal controller achieving
$M$ in closed-loop is the dynamical system giving $u_{t}^{lin}=u_{t}-u_{t}^{nl}$ as an output when fed by $e_{t}^{v}$. The command input $u_{t}^{lin}$
is the output of the extended PID controller \eqref{Klin}, which
can then be designed based on the data set \eqref{eq:data}.

Let $\tilde{\boldsymbol{u}}=(\tilde{u}_{1-L},\ldots,\tilde{u}_{0})$
be the input sequence of the data set and let $\tilde{\boldsymbol{u}}^{nl}=(\tilde{u}_{1-L}^{nl},\ldots,\tilde{u}_{0}^{nl})$
be obtained from the off-line filtering of $r_{t}^{v}$ and $\tilde{y}_{t}$,
$t=1-L,\ldots,0$, with the controller $K^{nl}$ derived in Section
\ref{sub:nl_des}.

Therefore, the control design problem can be turned into an identification
problem, where the optimal controller with the structure in \eqref{Klin}
is the one with parameter vector 
\begin{equation}
\theta=\arg\min_{\vartheta\in\mathbb{R}^{n_{\theta}}}\left\Vert \tilde{\boldsymbol{\delta u}}-\boldsymbol{u}^{lin}(\vartheta)\right\Vert _{2}^{2}\label{eq:theta0}
\end{equation}
where $\tilde{\boldsymbol{\delta u}}=\tilde{\boldsymbol{u}}-\tilde{\boldsymbol{u}}^{nl}$
and $\boldsymbol{u}^{lin}(\theta)=\left(u_{1-L}^{lin}(\theta),\ldots,u_{0}^{lin}(\theta)\right)$.
In \cite{campi2002virtual}, it is shown how the problem \eqref{eq:theta0}
can be solved by means of convex optimization.

\section{Closed-loop analysis}

\subsection{Stability analysis}

\label{sec:stab_an}

The feedback system of Figure \ref{ctr_sys_1} is described by 
\begin{equation}
\begin{array}{l}
y_{t+1}=g\left(\boldsymbol{y}_{t},\boldsymbol{u}_{t},\boldsymbol{\xi}_{t}\right)\\
u_{t}=u_{t}^{nl}+u_{t}^{lin}\\
u_{t}^{nl}=K^{nl}\left(r_{t+1},\boldsymbol{y}_{t},\boldsymbol{u}_{t-1}^{nl}\right)\\
u_{t}^{lin}=K^{lin}\left(\boldsymbol{r}_{t}-\boldsymbol{y}_{t},\boldsymbol{u}_{t-1}^{lin}\right)
\end{array}\label{eq:fb}
\end{equation}
where $K^{nl}$ and $K^{lin}$ are two functions, chosen to be Lipschitz
continuous, describing the nonlinear and linear controllers.

The assumptions required to guarantee the stability of this feedback
system are now introduced and commented.

\medskip{}

\begin{assumption}\label{ass:lip_b}The function $g$ in \eqref{ss_sys}
is Lipschitz continuous on $Y^{n}\times U^{n}\times\Xi^{n}$. Without
loss of generality, it is also assumed that $Y^{n}\times U^{n}\times\Xi^{n}$
contains the origin.$\qquad\blacksquare$ \end{assumption}\medskip{}

This assumption is mild, since most real-world dynamic systems are
described by functions that are Lipschitz continuous on a compact
set. Note anyway that all what presented in this paper can be easily
extended to the case where $g$ is the sum of a Lipschitz continuous
function plus a discontinuous but bounded function.

From Assumption \ref{ass:lip_b}, it follows that $g$ can be written
as
\[
g\left(\boldsymbol{y}_{t},\boldsymbol{u}_{t},\boldsymbol{\xi}_{t}\right)=g^{o}\left(\boldsymbol{y}_{t},\boldsymbol{u}_{t}\right)+g_{t}^{\xi}\boldsymbol{\xi}_{t}
\]
where $g^{o}\left(\boldsymbol{y}_{t},\boldsymbol{u}_{t}\right)\doteq g\left(\boldsymbol{y}_{t},\boldsymbol{u}_{t},\mathbf{0}\right)$,
$g_{t}^{\xi}\in\mathbb{R}^{1\times n}:\left\Vert g_{t}^{\xi}\right\Vert _{\infty}\leq\gamma_{\xi}$
for some $\gamma_{\xi}<\infty$. Assumption \ref{ass:lip_b} implies
that the residue function 
\[
\varDelta\left(\boldsymbol{y}_{t},\boldsymbol{u}_{t}\right)\doteq g^{o}\left(\boldsymbol{y}_{t},\boldsymbol{u}_{t}\right)-f\left(\boldsymbol{y}_{t},\boldsymbol{u}_{t}\right)
\]
 is Lipschitz continuous on $Y^{n}\times U^{n}\times\Xi^{n}$. In
particular, a finite and non-negative constant $\gamma_{y}$ exists,
such that
\[
\left\Vert \varDelta\left(y,u\right)-\varDelta\left(y',u,\right)\right\Vert \leq\gamma_{y}\left\Vert y-y'\right\Vert _{\infty}
\]
for all $y,y'\in Y^{n}$. \medskip{}

\begin{assumption}\label{ass:lip_delta}$\gamma_{y}\leq1$.$\qquad\blacksquare$
\end{assumption}\medskip{}

The meaning of this assumption is clear: it requires that $f$ describes
accurately the variability of $g$ with respect to $\boldsymbol{y}_{t}$.
\medskip{}

In order to introduce the next assumption, the following stability
notion is needed.

\medskip{}

\begin{definition}A nonlinear (possibly time-varying) system with
input $u_{t}$, output $y_{t}$ and noise $\xi_{t}$ is \emph{finite-gain
$\ell_{\infty}$ stable} on $\left(\mathcal{Y}^{0},\mathcal{U},\varXi\right)$
if finite and non-negative constants $\Gamma_{u}$, $\Gamma_{\xi}$
and $\Lambda$ exist such that
\[
\left\Vert \boldsymbol{y}\right\Vert _{\infty}\leq\Gamma_{u}\left\Vert \boldsymbol{u}\right\Vert _{\infty}+\Gamma_{\xi}\left\Vert \boldsymbol{\xi}\right\Vert _{\infty}+\Lambda
\]
for any $(\boldsymbol{y}_{0},\boldsymbol{u},\boldsymbol{\xi})\in\mathcal{Y}^{0}\times\mathcal{U}\times\varXi$,
where $\boldsymbol{u}=(u_{1},u_{2},\ldots)$, $\boldsymbol{\xi}=(\xi_{1},\xi_{2},\ldots)$
and $\boldsymbol{y}=(y_{1},y_{2},\ldots)$.$\qquad\blacksquare$ \end{definition}\medskip{}

Note that this finite-gain\emph{ }stability definition is more general
than the standard one, which corresponds to the case $\mathcal{U}=\ell_{\infty}$
and $\varXi=\ell_{\infty}$, see e.g. \cite{Khalil96}.

Now, consider that the difference equation \eqref{eq:model}, where
$u_{t}$ is given in \eqref{eq:fb}, defines a dynamical system with
inputs $y_{t}$ and $r_{t+1}$, and output $\hat{y}_{t+1}$ ($u_{t}$,
$u_{t}^{nl}$ and $u_{t}^{lin}$ are internal variables). This system
is finite-gain $\ell_{\infty}$ stable on $\left(\ell_{\infty},\ell_{\infty},\ell_{\infty}\right)$:
\begin{equation}
\left\Vert \hat{\boldsymbol{y}}\right\Vert _{\infty}\leq\Gamma_{y}\left\Vert \boldsymbol{y}\right\Vert _{\infty}+\Gamma_{r}\left\Vert \boldsymbol{r}\right\Vert _{\infty}+\Lambda_{f}\label{eq:ass_f}
\end{equation}
with $\Gamma_{y},\Gamma_{r},\Lambda_{f}<\infty$. In fact, the system
is formed by the cascade connection of the controller and the model
\eqref{eq:model}. The controller provides a command input $u_{t}$
bounded in the compact set $U$. The model is a static Lipschitz continuous
function of a regressor consisting in past values of $u_{t}$ and
$y_{t}$.\medskip{}

\begin{assumption}\label{ass:clf_stab}$\Gamma_{y}<1-\gamma_{y}$.$\qquad\blacksquare$\end{assumption}\medskip{}

This assumption is not restrictive: It is certainly satisfied if $\mu=0$
and the reference $\boldsymbol{r}=(r{}_{1},r_{2},\ldots)$ is a system
solution (i.e. $r_{t+1}$ is in the range of $f\left(\boldsymbol{y}_{t},\cdot\right)$
for all $t$). Indeed, in this case, $\hat{y}_{t+1}=r_{t+1}$, $\forall t$,
since $K^{nl}$ performs an exact inversion of the model, see \eqref{eq:opt2})
($K^{lin}$ gives a null input signal). This implies that $\Gamma_{y}=0$,
$\Gamma_{r}=1$ and $\Lambda_{f}=0$. Hence, if a ``not too large''
$\mu$ is chosen and the reference is ``not too far'' from a system
solution, supposing that inequality \eqref{eq:ass_f} holds with a
``small'' $\Gamma_{y}$ is completely reasonable. The meaning of Assumption
\ref{ass:clf_stab} is that, in order to guarantee closed-loop stability,
the controller must perform an effective right-inversion of the system
and this inversion should depend as less as possible on the current
working point $\boldsymbol{y}_{t}$. Note that the bound \eqref{eq:ass_f}
implies that, if the model \eqref{eq:model} is exact, the designed
controller stabilizes the closed loop system (a direct consequence
of Theorem \ref{thm:stab} below).

From Assumption \ref{ass:clf_stab} it follows that the system defined
by the difference equation $\hat{e}_{t}=r_{t}-f\left(\boldsymbol{y}_{t-1},u_{t-1}\right)$,
where $u_{t}$ is given in \eqref{eq:fb}, is finite-gain $\ell_{\infty}$
stable on $\left(\mathbf{0},\mathcal{Y},\mathcal{Y}\right)$:
\begin{equation}
\left\Vert \hat{\boldsymbol{e}}\right\Vert _{\infty}\leq\Gamma_{y}\left\Vert \boldsymbol{y}\right\Vert _{\infty}+\Gamma_{s}\left\Vert \boldsymbol{r}\right\Vert _{\infty}+\Lambda_{e}\label{eq:bbe}
\end{equation}
with $\Gamma_{y}<1-\gamma_{y}$ and $\Gamma_{s},\Lambda_{e}<\infty$.
As discussed above, in ``reasonable'' working conditions, $\hat{y}_{t}\cong r_{t}$,
implying that $\Gamma_{y}\cong0$ and $\Gamma_{s}\cong0$. \medskip{}

Closed-loop stability of the system \eqref{eq:fb} is stated by the
following result, which also provides a bound on the tracking error.
\medskip{}

\begin{theorem}\label{thm:stab}Let Assumptions \ref{ass:lip_b}-\ref{ass:clf_stab}
hold. Then:\\
(i) For any initial condition $\boldsymbol{y}_{0}\in\mathcal{Y}^{0}$,
the feedback system \eqref{eq:fb} is finite-gain $\ell_{\infty}$
stable on $\left(\mathcal{Y}^{0},\mathcal{Y}^{S},\varXi\right)$:
\[
\left\Vert \boldsymbol{y}\right\Vert _{\infty}\leq\frac{1}{1-\Gamma_{y}-\gamma_{y}}\left(\Gamma_{r}\left\Vert \boldsymbol{r}\right\Vert _{\infty}+\gamma_{\xi}\left\Vert \boldsymbol{\xi}\right\Vert _{\infty}+\Lambda_{g}\right)
\]
where $\Lambda_{g}\doteq\Lambda_{f}+\max_{u\in U^{n}}\left\Vert \varDelta\left(\mathbf{0},u\right)\right\Vert _{\infty}<\infty$
and
\[
\mathcal{Y}^{S}\doteq\left\{ \boldsymbol{r}\in\mathcal{Y}\left(\boldsymbol{y}_{0}\right):y_{t}\in Y,\forall t,\forall\boldsymbol{\xi}\in\varXi\right\} .
\]
(ii) The tracking error $\boldsymbol{e}\doteq\boldsymbol{r}-\boldsymbol{y}$
is bounded as
\[
\left\Vert \boldsymbol{e}\right\Vert _{\infty}\leq\frac{1}{1-\Gamma_{y}}\left(\Gamma_{er}\left\Vert \boldsymbol{r}\right\Vert _{\infty}+\gamma_{\xi}\left\Vert \boldsymbol{\xi}\right\Vert _{\infty}+\Lambda_{e}+\left\Vert \Delta\right\Vert _{\infty}\right)
\]
where $\Gamma_{er}\doteq\Gamma_{y}+\Gamma_{s}$ and $\left\Vert \Delta\right\Vert _{\infty}$
is the functional $L_{\infty}$ norm of $\Delta$, evaluated over
$Y^{n}\times U^{n}$.\end{theorem}\medskip{}

\textbf{Proof.} (i) The feedback system of Figure \ref{ctr_sys_1}
is described by 
\begin{equation}
y_{t+1}=g\left(\boldsymbol{y}_{t},\boldsymbol{u}_{t},\boldsymbol{\xi}_{t}\right)=\hat{y}_{t+1}+\delta y_{t}\label{eq:pr1_1}
\end{equation}
where 
\[
\begin{array}[t]{l}
\hat{y}_{t+1}=f\left(\boldsymbol{y}_{t},\boldsymbol{u}_{t}\right)\\
\delta y_{t}=\varDelta\left(\boldsymbol{y}_{t},\boldsymbol{u}_{t}\right)+g_{t}^{\xi}\boldsymbol{\xi}_{t}
\end{array}
\]
and $u_{t}$ is given by \eqref{eq:fb}.

From \eqref{eq:pr1_1} and Assumption \ref{ass:clf_stab}, the following
inequalities hold:
\begin{equation}
\begin{array}{l}
\left\Vert y_{t+1}\right\Vert _{\infty}\leq\left\Vert \hat{y}_{t+1}\right\Vert _{\infty}+\left\Vert \delta y_{t}\right\Vert _{\infty}\\
\leq\Gamma_{r}\left\Vert \boldsymbol{r}\right\Vert _{\infty}+\Gamma_{y}\left\Vert \boldsymbol{y}\right\Vert _{\infty}+\Lambda_{f}+\left\Vert \delta y_{t}\right\Vert _{\infty}.
\end{array}\label{eq:pr1_2}
\end{equation}
In order to derive a bound on $\left\Vert \delta y_{t}\right\Vert _{\infty}$,
consider that,
\[
\begin{array}[t]{c}
\left\Vert \varDelta\left(\boldsymbol{y}_{t},\boldsymbol{u}_{t}\right)\right\Vert _{\infty}-\left\Vert \varDelta\left(\mathbf{0},\boldsymbol{u}_{t}\right)\right\Vert _{\infty}\\
\leq\left\Vert \varDelta\left(\boldsymbol{y}_{t},\boldsymbol{u}_{t}\right)-\varDelta\left(\mathbf{0},\boldsymbol{u}_{t}\right)\right\Vert _{\infty}\leq\gamma_{y}\left\Vert \boldsymbol{y}_{t}\right\Vert _{\infty}.
\end{array}
\]
This inequality is due to Assumption \ref{ass:lip_delta} and holds
for any $\boldsymbol{r}\in\mathcal{Y}^{S}$. Indeed, $\mathcal{Y}^{S}$
is the set of all reference sequences for which the system output
remains in the domain where $\Delta$ is Lipschitz continuous with
respect to $\boldsymbol{y}_{t}$ with constant $\gamma_{y}$ . 

It follows that
\begin{equation}
\begin{array}{c}
\left\Vert \delta y_{t}\right\Vert _{\infty}\leq\left\Vert \varDelta\left(\boldsymbol{y}_{t},\boldsymbol{u}_{t}\right)\right\Vert _{\infty}+\gamma_{\xi}\left\Vert \boldsymbol{\xi}_{t}\right\Vert _{\infty}\\
\leq\gamma_{y}\left\Vert \boldsymbol{y}_{t}\right\Vert _{\infty}+\gamma_{\xi}\left\Vert \boldsymbol{\xi}_{t}\right\Vert _{\infty}+\bar{\Delta}\\
\leq\gamma_{y}\left\Vert \boldsymbol{y}\right\Vert _{\infty}+\gamma_{\xi}\left\Vert \boldsymbol{\xi}\right\Vert _{\infty}+\bar{\Delta}
\end{array}\label{eq:pr1_3}
\end{equation}
where 
\[
\bar{\Delta}\doteq\max_{u\in U^{n}}\left\Vert \varDelta\left(\mathbf{0},u\right)\right\Vert _{\infty}.
\]
Note that $\bar{\Delta}<\infty$ since $\Delta$ is Lipschitz continuous
and $U$ is a compact set, implying that $\Lambda_{g}<\infty$.

From \eqref{eq:pr1_2} and \eqref{eq:pr1_3}, we obtain:
\[
\begin{array}[t]{c}
\left\Vert y_{t+1}\right\Vert _{\infty}\leq\Gamma_{r}\left\Vert \boldsymbol{r}\right\Vert _{\infty}+\Gamma_{y}\left\Vert \boldsymbol{y}\right\Vert _{\infty}+\Lambda_{f}\\
+\gamma_{y}\left\Vert \boldsymbol{y}\right\Vert _{\infty}+\gamma_{\xi}\left\Vert \boldsymbol{\xi}\right\Vert _{\infty}+\bar{\Delta}.
\end{array}
\]
Since this inequality holds for all $t$, we have that
\[
\begin{array}[t]{c}
\left\Vert \boldsymbol{y}\right\Vert _{\infty}\leq\Gamma_{r}\left\Vert \boldsymbol{r}\right\Vert _{\infty}+\Gamma_{y}\left\Vert \boldsymbol{y}\right\Vert _{\infty}+\Lambda_{f}\\
+\gamma_{y}\left\Vert \boldsymbol{y}\right\Vert _{\infty}+\gamma_{\xi}\left\Vert \boldsymbol{\xi}\right\Vert _{\infty}+\bar{\Delta},
\end{array}
\]
which yields the following bound:
\[
\left\Vert \boldsymbol{y}\right\Vert _{\infty}\leq\frac{1}{1-\Gamma_{y}-\gamma_{y}}\left(\Gamma_{r}\left\Vert \boldsymbol{r}\right\Vert _{\infty}+\gamma_{\xi}\left\Vert \boldsymbol{\xi}\right\Vert _{\infty}+\Lambda_{f}+\bar{\Delta}\right)
\]
where it has been considered that, by Assumptions \ref{ass:clf_stab}
and \ref{ass:lip_delta}, $\Gamma_{y}+\gamma_{y}<1$.

(ii) From \eqref{eq:pr1_1}, we have that
\[
e_{t}\doteq r_{t}-y_{t}=r_{t}-\hat{y}_{t}-\delta y_{t}.
\]
Then,
\[
\left\Vert e_{t}\right\Vert _{\infty}\leq\left\Vert r_{t}-\hat{y}_{t}\right\Vert _{\infty}+\left\Vert \delta y_{t-1}\right\Vert _{\infty}.
\]
The term $\left\Vert r_{t}-\hat{y}_{t}\right\Vert _{\infty}=\left\Vert \hat{e}_{t}\right\Vert _{\infty}$
is bounded according to \eqref{eq:bbe}. The following bound on the
term $\left\Vert \delta y_{t-1}\right\Vert _{\infty}$ is considered:
\[
\left\Vert \delta y_{t-1}\right\Vert _{\infty}\leq\left\Vert \Delta\right\Vert _{\infty}+\gamma_{\xi}\left\Vert \boldsymbol{\xi}\right\Vert _{\infty},
\]
which holds for any $\boldsymbol{r}\in\mathcal{Y}^{S}$. Thus,
\[
\begin{array}[t]{c}
\left\Vert \boldsymbol{e}\right\Vert _{\infty}\leq\Gamma_{y}\left\Vert \boldsymbol{y}\right\Vert _{\infty}+\Gamma_{s}\left\Vert \boldsymbol{r}\right\Vert _{\infty}\\
+\Lambda_{e}+\left\Vert \Delta\right\Vert _{\infty}+\gamma_{\xi}\left\Vert \boldsymbol{\xi}\right\Vert _{\infty}\\
\\
\leq\Gamma_{y}\left\Vert \boldsymbol{e}\right\Vert _{\infty}+\Gamma_{y}\left\Vert \boldsymbol{r}\right\Vert _{\infty}+\Gamma_{s}\left\Vert \boldsymbol{r}\right\Vert _{\infty}\\
+\Lambda_{e}+\left\Vert \Delta\right\Vert _{\infty}+\gamma_{\xi}\left\Vert \boldsymbol{\xi}\right\Vert _{\infty}.
\end{array}
\]
The claim follows.$\qquad\blacksquare$\medskip{}

Theorem \ref{thm:stab} can be interpreted as follows. Two main conditions
are sufficient to guarantee closed-loop stability. First, the model
must describe accurately the model rate of variation with respect
to $\boldsymbol{y}_{t}$ (i.e., the constant $\gamma_{y}$ in Assumption
\ref{ass:lip_delta} must be small). Second, the controller has to
perform an effective inversion of the model (Assumption \ref{ass:clf_stab}).
These conditions allow for closed-loop stability and lead to the tracking
error bound given in Theorem \ref{thm:stab}. It can be noted that
this error is reduced if the model provides a ``small'' prediction
error ($\left\Vert \Delta\right\Vert _{\infty}$ is a measure of the
prediction error on the whole model domain). In summary, the model
should thus satisfy two requirements: it must be accurate in describing
the dependence on $\boldsymbol{y}_{t}$ and, at the same time, in
reproducing the system output. Note that, in the proposed control
scheme, the model does not work in simulation but in prediction.

These results hold for any reference $\boldsymbol{r}\in\mathcal{Y}^{S}$,
where $\mathcal{Y}^{S}$ is the set of all sequences of interest for
which the system output remains in the domain where $\Delta$ is Lipschitz
continuous with respect to $\boldsymbol{y}_{t}$, with fixed constant
$\gamma_{y}$. Clearly, this domain must be well explored by the data
\eqref{eq:data}, in order to ensure the accuracy properties described
above.

A reliable indication for generating suitable references can be the
following: a reference signal should be a solution (or an approximate
solution) of the system to control, i.e. a signal $\boldsymbol{r}=(r_{1},r_{2},\ldots)$
for which, at each time $t$, a $u_{t}$ exists giving $y_{t+1}=g\left(\boldsymbol{y}_{t},u_{t},u_{t-1},\ldots,u_{t-n+1},\boldsymbol{\xi}_{t}\right)\cong r_{t+1}$.
More in general, the reference trajectory must be compatible with
the physical properties of the system to control. For instance,
in a second order mechanical system, the two states are typically
a position and a velocity. Thus, the position reference can be generated
as a sequence of values ranging in the physical domain of this variable
with ``not too high'' variations (no other particular indications
are required here). The velocity reference can obviously be generated
as the derivative of the position reference. Note anyway that reference
design is a well-known open problem which arises for most nonlinear
identification and control methods.\\

\begin{remark}The stability analysis developed in the present paper
is substantially different from the one in \cite{NoFaMiAUT13}. Indeed,
no model is identified in \cite{NoFaMiAUT13}. The controller (directly
designed from data) is seen as an approximation of some unknown ideal
controller. The stability conditions depend on the quality of this
approximation. In the present paper, no ideal controllers are assumed.
The stability conditions are related to the quality of the identified
model.$\qquad\blacksquare$

\end{remark}

\subsection{Properties of the linear controller}

\label{sub:lin_an}

The stability analysis of Section \ref{sec:stab_an} has been carried
out considering the nonlinear and linear controllers together, as
a unique block. The importance of the nonlinear controller is evident
from the above results and discussions. An analysis is now carried
out, showing that the linear controller is fundamental to further
increase the tracking precision and robustness of the feedback system.

\begin{figure}[h!tb]
\centering \includegraphics[width=0.5\columnwidth]{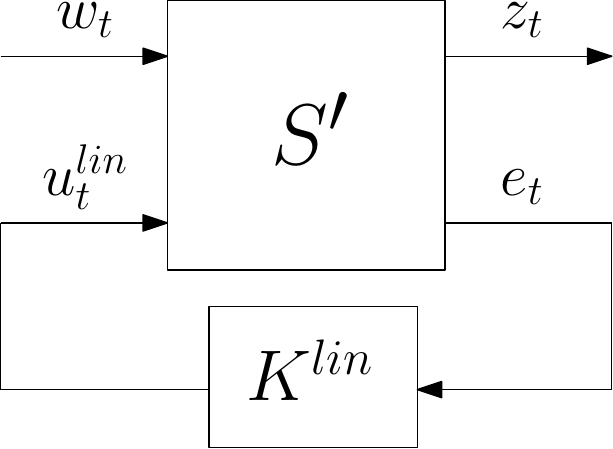} \protect\caption{LFT representation of the nonlinear control system with the linear
feedback.}

\label{fig:lft} 
\end{figure}

First of all, let us introduce the following assumption on the closed-loop
system with the nonlinear controller.\medskip{}

\begin{assumption}\label{ass:lin} Let $S'$ be the system formed
by the feedback interconnection of $S$ and $K^{nl}$, having inputs
$u_{t}^{lin}$, $r_{t}$ and $\xi_{t}$ and output $e_{t}=r_{t}-y_{t}$.
The action $u_{t}^{lin}$ of the linear controller $K^{lin}$ and $\xi_t$ are
sufficiently small, so that $S'$ is assumed to be characterized by an LTI (Linear
Time Invariant) behavior. $\qquad\blacksquare$ \end{assumption}\medskip{}

This assumption is justified by the fact that the nonlinear controller
$K^{nl}$, if correctly designed, brings the system close to a desired
trajectory. It is thus reasonable supposing that the behavior of the
system in a sufficiently small neighborhood of the trajectory is linear.
Note also that the assumption is quite mild, as no specific dynamic
description is required for $S'$. The variations of the signals from
a given operating trajectory are simply required to be ``small''.

Under Assumption \ref{ass:lin}, the overall control system of Figure
\ref{ctr_sys_1} can be represented in an LFT (Linear Fractional Transformation)
fashion as in Figure \ref{fig:lft}, where 
\[
z_{t}=e_{t},\ w_{t}=[r_{t},\ \xi_{t}]^{T},
\]
\[
\begin{bmatrix}z_{t}\\
e_{t}
\end{bmatrix}=\begin{bmatrix}G_{11} & G_{12}\\
G_{21} & G_{22}
\end{bmatrix}\begin{bmatrix}w_{t}\\
u_{t}^{lin}
\end{bmatrix}
\]
and the $G_{ij}$'s are unknown transfer functions for all $i,j$.
Notice that these transfer functions may even be unstable. Instead,
with $K^{lin}$, not only the overall system is stable for Theorem
\ref{thm:stab}, but the steady-state performance is definitely enhanced,
as illustrated by the following result.\medskip{}

\begin{theorem}\label{thm:linear} Under Assumption \ref{ass:lin},
the System $S'$ with a linear controller $K^{lin}$ of type \eqref{Klin}
designed according to the D$^{2}$-IBC approach, is such that\\
 (i) the steady state tracking error for a reference step excitation
is zero;\\
 (ii) any constant disturbance $\xi_{t}$ gives zero steady-state
contribution to $e_{t}$. $\qquad\blacksquare$ \end{theorem}\medskip{}

\textbf{Proof.} From robust control theory \cite{doyle1992feedback},
the transfer function between $w_{t}$ and $z_{t}$ in the scheme
of Fig. \ref{fig:lft} is 
\[
T_{zw}=G_{11}+G_{12}K^{lin}(I-G_{22}K^{lin})^{-1}G_{21}.
\]
Since, in our case, $z_{t}=e_{t}$, then $G_{11}=G_{21}$, $G_{12}=G_{22}$
and 
\[
T_{zw}=G_{11}+G_{22}K^{lin}(I-G_{22}K^{lin})^{-1}G_{11}.
\]
(i) Now consider only the contribution of $r_{t}$ on $z_{t}$ and
let $G_{11}=[G_{11}^{r}\ G_{11}^{\xi}]$. The transfer function between
$r_{t}$ and $z_{t}=e_{t}$ is the first element of $T_{zw}=[T_{zr}\ T_{z\xi}]$,
that is 
\[
T_{zr}=G_{11}^{r}+\frac{G_{22}G_{11}^{r}K^{lin}}{1-G_{22}K^{lin}}=\frac{1}{1-G_{22}K^{lin}}.
\]
Since the controller $K^{lin}$ of type \eqref{Klin} contains an
integrator and the overall controller is such that the final system
is stable (i.e. the numerator of $1-G_{22}K^{lin}$ must have only
stable roots), $T_{zr}$ turns out to have a derivative action which
gives zero steady-state response to any reference step for all $G_{11}$
and $G_{22}$.\\
 (ii) The proof of this point comes straightforwardly from the previous
one, by noting that also 
\[
T_{z\xi}=G_{11}^{\xi}+\frac{G_{22}G_{11}^{\xi}K^{lin}}{1-G_{22}K^{lin}}=\frac{1}{1-G_{22}K^{lin}}
\]
contains a derivator. $\qquad\blacksquare$\medskip{}

Notice that, without $K^{lin}$, the relationship between $w_{t}$
and the tracking error $e_{t}$ can in principle be anything. Intuitively,
if the nonlinear inversion operated by $K^{nl}$ is accurate enough,
the asymptotic effect of the external disturbances on the error will
be small. However, Theorem \ref{thm:linear} shows that only the linear
controller can guarantee that such an effect is exactly zero. \medskip{}

\begin{remark}
Notice that the above results may hold also when $S'$ is linear but not time-invariant, provided that the linear relationships can be written - for any $i$, $j$ - as $G_{ij}=\tilde{G}_{ij}+\Delta G$, where $\tilde{G}_{ij}$ is a time-invariant nominal term and $\Delta G$ is another time-invariant term upper-bounding the time-varying dynamics, namely a system with a bound $\delta_G<\infty$ such that $\left\|\Delta G\right\|_{\infty}\leq \delta G$.
$\qquad\blacksquare$

\end{remark}

\bibliographystyle{IEEEtran}
\bibliography{lettnos_journals,lettnos_conferences,lettaltr,simo_biblio}

\end{document}